\documentclass[conference]{IEEEtran}
\IEEEoverridecommandlockouts

\usepackage{cite}
\usepackage[dvipdfmx]{graphicx}
\usepackage{tabularx} 
\usepackage{booktabs} 
\usepackage{latexsym} 
\usepackage{verbatim}
\usepackage{moreverb}
\usepackage{makeidx} 
\usepackage{float}
\usepackage{here}
\usepackage{multirow}
\usepackage{graphics}
\usepackage{amssymb}
\usepackage{amsmath}
\usepackage{comment}
\usepackage{url}
\usepackage{xcolor}
\usepackage{graphicx}
\usepackage{algorithm}
\usepackage{algorithmic}
\usepackage{multirow}
\usepackage{bbding}
\usepackage{pifont}
\usepackage{wasysym}
\usepackage{amssymb}
\usepackage{listings}
\usepackage{microtype}
\usepackage{seqsplit}
\usepackage{color}
\usepackage{hyperref}

\newif\ifdraft
\draftfalse

\ifdraft
\newcommand{\cone}[1]{\color{red} #1 \color{black}}
\newcommand{\ctwo}[1]{\color{green} #1 \color{black}}
\newcommand{\cthree}[1]{\color{blue} #1 \color{black}}
\else
\newcommand{\cone}[1]{#1}
\newcommand{\ctwo}[1]{#1}
\newcommand{\cthree}[1]{#1}
\fi

\def\BibTeX{{\rm B\kern-.05em{\sc i\kern-.025em b}\kern-.08em
    T\kern-.1667em\lower.7ex\hbox{E}\kern-.125emX}}

\begin{document}

\title{Model-based Development for\\Autonomous Driving Systems Considering\\Event-driven and Timer-driven Nodes}

\author{\IEEEauthorblockN{Kenshin Obi}
\IEEEauthorblockA{\textit{Graduate School of}\\
\textit{Science and Engineering}\\
\textit{Saitama University}}
\and
\IEEEauthorblockN{Ryo Yoshinaka}
\IEEEauthorblockA{\textit{Graduate School of}\\
\textit{Science and Engineering}\\
\textit{Saitama University}}
\and
\IEEEauthorblockN{Hiroshi Fujimoto}
\IEEEauthorblockA{\textit{Software Division}\\
\textit{eSOL Co., Ltd}}
\and
\IEEEauthorblockN{Takuya Azumi}
\IEEEauthorblockA{\textit{Graduate School of}\\
\textit{Science and Engineering}\\
\textit{Saitama University}}
}

\maketitle
\IEEEpubid{\makebox[\columnwidth]{© 2024 IEEE. Personal use of this material is permitted.
DOI: 10.1109/SEAA64295.2024.00017\hfill}}
\IEEEpubidadjcol


\begin{abstract}
In recent years, the complexity and scale of embedded systems, especially in the rapidly developing field of autonomous driving systems, have increased significantly. This has led to the adoption of software and hardware approaches such as Robot Operating System (ROS)~2 and multi-core processors. Traditional manual program parallelization faces challenges, including maintaining data integrity and avoiding concurrency issues such as deadlocks. While model-based development (MBD) automates this process, it encounters difficulties with the integration of modern frameworks such as ROS~2 in multi-input scenarios.
This paper proposes an MBD framework to overcome these issues, categorizing ROS~2-compatible Simulink models into event-driven and timer-driven types for targeted parallelization. As a result, it extends the conventional parallelization by MBD and supports ROS~2-based models with multiple inputs.
The evaluation results show that after applying parallelization with the proposed framework, all patterns show a reduction in execution time, confirming the effectiveness of parallelization.

\end{abstract}


\section{Introduction}
In recent years, the complexity and scale of development and design of embedded systems, including autonomous driving systems~\cite{Autoware}, has increased. This trend has encouraged the adoption of advanced software frameworks and hardware architectures. For example, ROS~2~\cite{Exploring}, the second generation of the Robot Operating System~\cite{Exploring, ROS}, is increasingly being used as middleware for autonomous driving systems. Concurrently, the use of multi-core processors is rapidly expanding in areas that require demanding real-time processing, including such as autonomous driving systems.

To implement parallel processing in a multi-core environment, it is essential to design in consideration of task allocation to processor cores, synchronous processing and avoidance of memory access contention. One approach to design is automatic parallelisation using model-based development (MBD), an approach in which a system is designed using mathematical models and simulation before the actual prototype is created. Model-Based Parallelizer (MBP) used in MBD performs parallelization by considering the dependencies and specific rules of MATLAB/Simulink~\cite{MATLAB} blocks, output C code suitable for multi/many-core processors. This approach contributes to shortening development time and improving the reliability of embedded systems. However, this approach still faces open issues, particularly concerning compatibility with ROS~2.

\IEEEpubidadjcol

The main reason why auto-parallelization using MBD has not yet been adapted to ROS~2 is that each technology was developed for different purposes; ROS~2 focuses on inter-node communication and collaboration. Traditional auto-parallelization using MBD, however, is geared toward optimizing isolated nodes or systems without the need for complex inter-system coordination. Against this background, adapting MBD-based parallelization methods to ROS~2 requires support for different processing timings for multiple inputs. In this paper, we define ``event-driven'' as a pattern that starts processing immediately based on input data, and ``timer-driven'' as a pattern that starts processing at regular intervals. It is essential to consider the processing timing for each of the event-driven and timer-driven types, and to parallelize them appropriately.

This paper proposes a model-based development framework that can handle both event-driven and timer-driven processing. The framework is based on C code parallelized by a model-based parallelization tool and incorporates ROS~2 input/output interfaces to extend functionality. The framework includes specific transformations to adapt to event-driven and timer-driven types. In addition, parallelized ROS~2-based C++ code is automatically generated with appropriate conversion to event-driven and timer-driven types. This approach is expected to meet real-time requirements and reduce development time.

The main contributions of this paper are as follows:
\begin{itemize}
  \item To enable parallelization of Simulink models through timer-driven processing.
  \item To enable parallelization of Simulink models through event-driven processing.
  \item To facilitate the execution of parallelized code on multi-core/many-core processors.
\end{itemize}

The remainder of this paper is organized as follows. Section~\ref{chapter:system model} describes the system model of MBP, ROS 2, and Many-core Processors, Section~\ref{chapter:proposed framework} describes the proposed framework. Section~\ref{chapter:evaluation} presents the evaluation and analysis, Section~\ref{chapter:related work} discusses related work. Section~\ref{chapter:conclusion} presents the conclusions and directions for future work.

\section{System Model}
\label{chapter:system model}
\begin{figure}[tb]
  \begin{center}
    \includegraphics[width=0.9\linewidth]{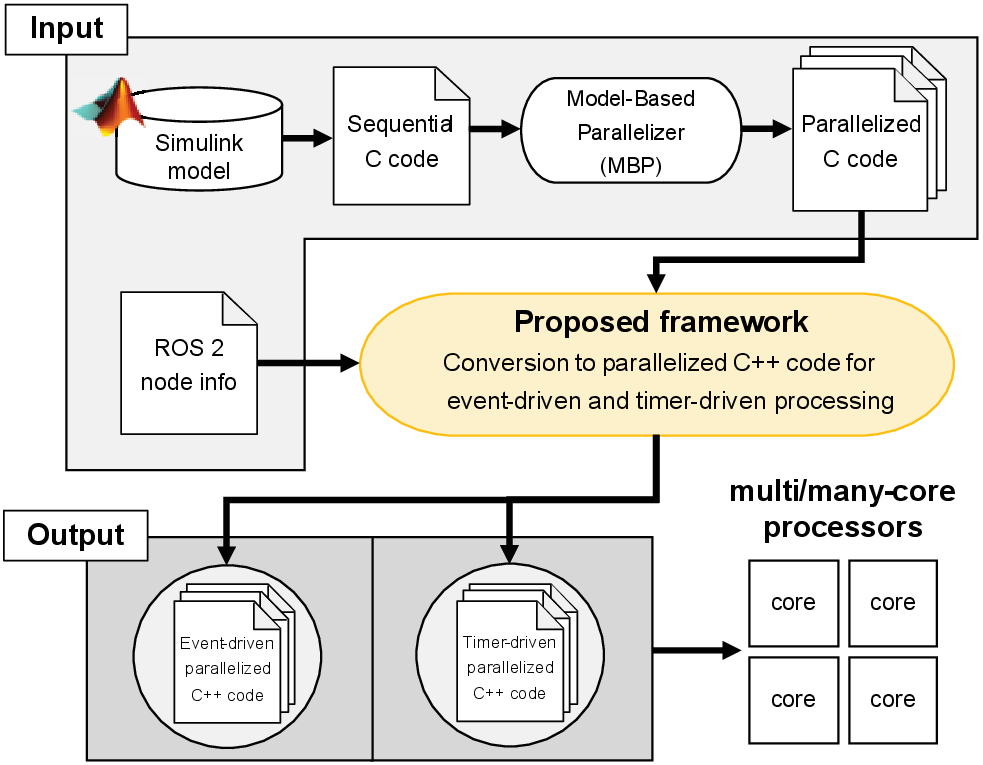}
    \caption{Overview of system model.}
    \label{fig:systemmodel}
  \end{center}
\end{figure}

  \begin{figure}[tb]
    \centering
    \includegraphics[width=1.0\linewidth]{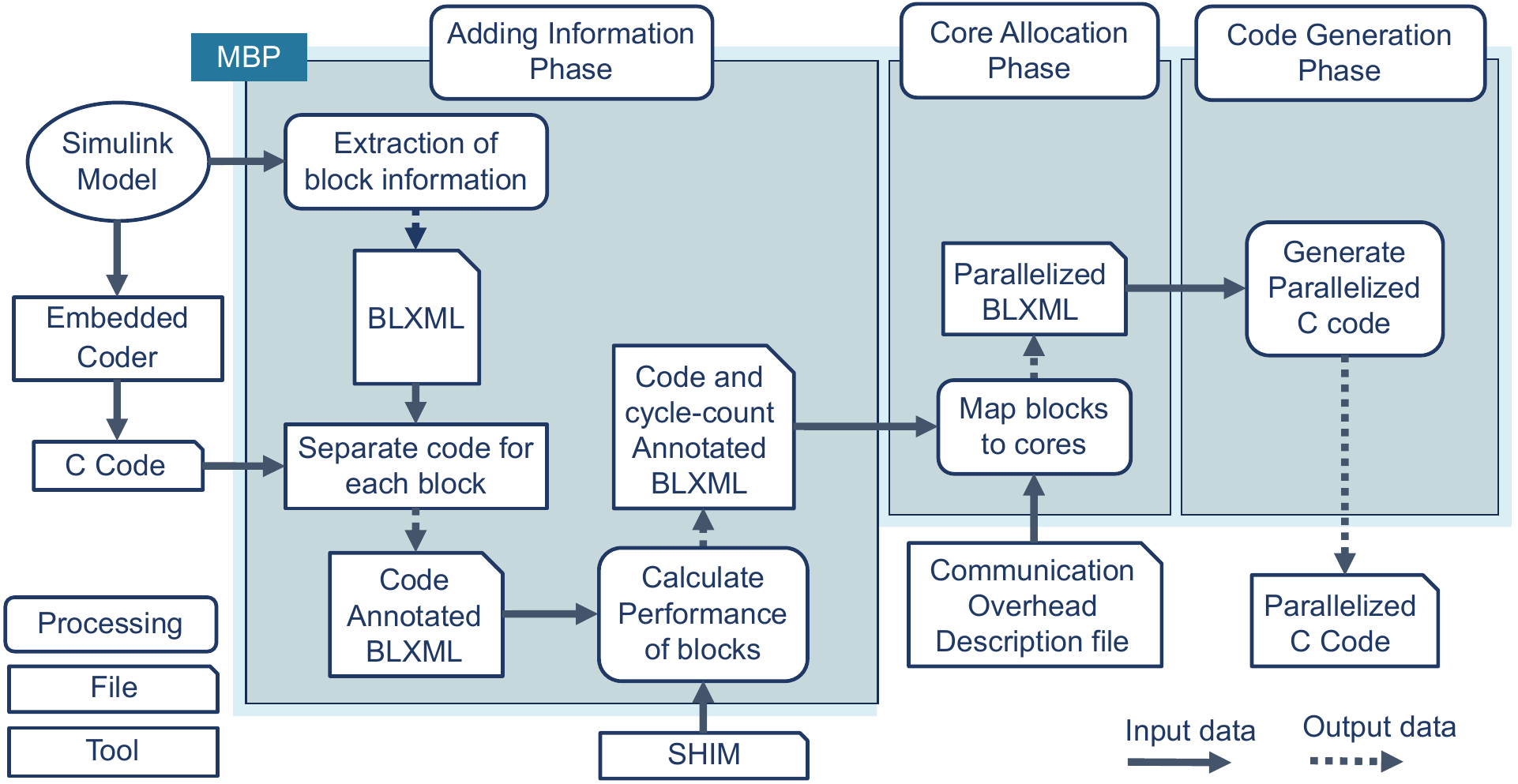}
    \caption{System model of model-based parallelizer (MBP).}
    \label{fig:MBP}
  \end{figure}

This section describes the system model as depicted in Fig.~\ref{fig:systemmodel}. The proposed framework is designed to convert Simulink models into parallelized C++ code for use in ROS~2 environments, supporting both event-driven and timer-driven nodes. First, the sequential C code generated from the Simulink model is used to generate parallelized code using MBP automatically. The proposed framework then performs transformations based on this parallelized C code and node information to generate parallelized C++ code. This code utilizes multi/many-core processors for parallel execution. First, Section~\ref{section:mbp} details MBP, a parallelization tool in model-based development. Next, Section~\ref{section:ros2} provides a basic explanation of ROS~2, and Section~\ref{section:many-core} discusses many-core processors used in this paper.

\subsection{MBP}
\label{section:mbp}
      
MBP can generate parallelized C code automatically from Simulink models, as shown in Fig.~\ref{fig:MBP}, improving development efficiency greatly. The MBP process consists of the following three phases.
    
\subsubsection{Information Adding Phase}

MBP generates BLXML (block structure information) from Simulink models. BLXML is an XML file that describes structural information such as block names, types, and connections of Simulink models. C code for a single core is generated with Embedded Coder~\cite{EmbededdCoder}. The C code is then combined with BLXML to generate BLXML with code annotations. In addition, MBP uses SHIM (Software-Hardware Interface for Multi-Many-Core) to calculate the performance of each block. SHIM is an abstract description of hardware information such as processor and memory, and by using SHIM, MBP can be applied easily to various hardware. Finally, MBP combines the performance information into BLXML, and generates BLXML with annotated code and cycle counts.

\subsubsection{Core Allocation Phase}

MBP maps blocks to processor cores to produce parallelized BLXML. Three conditions must be met for proper core allocation. First, the load balances of processor cores should be as similar to each other as possible. Second, the communication overhead should be as small as possible. Finally, the deadline should be kept.

\subsubsection{Code Generation Phase}
  
MBP reconstructs the C code for each block and generates a parallelized C code. According to the parallelized BLXML, which includes core allocations, MBP redesigns the C code in the parallelized BLXML for each allocated core.
  
\subsection{ROS 2}
\label{section:ros2}
  \begin{figure}[tb]
    \centering
    \includegraphics[width=0.9\linewidth]{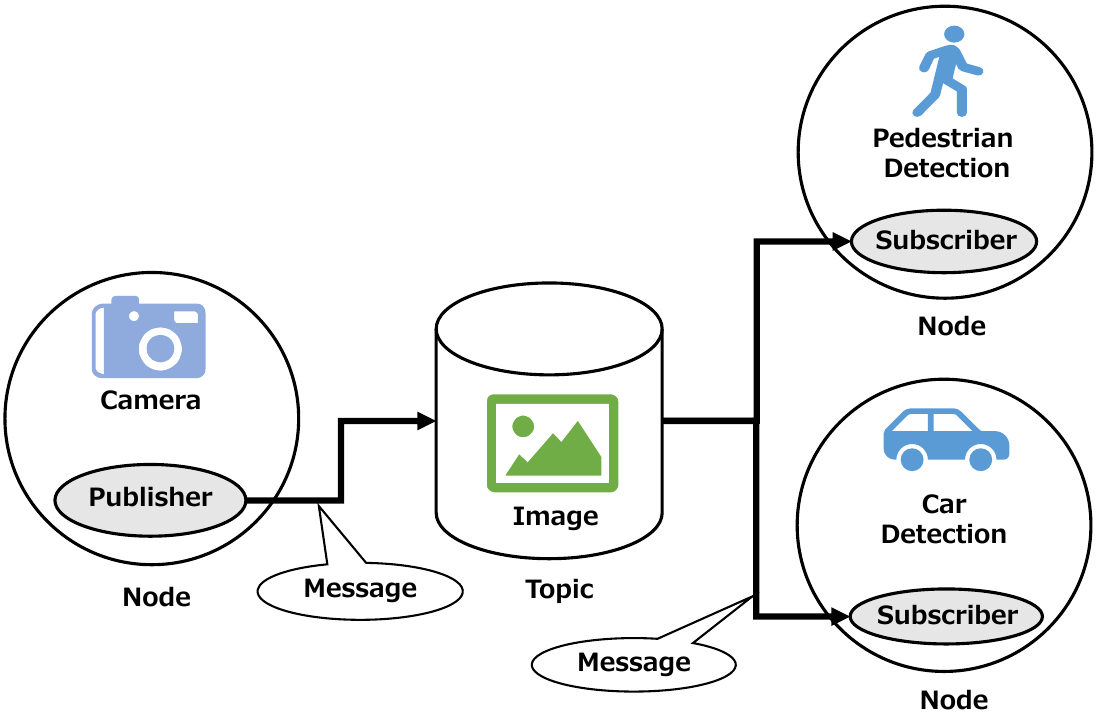}
    \caption{Publish/Subscribe model in ROS 2.}
    \label{fig:pubsub_model}
  \end{figure}
  
ROS~2, a general-purpose framework widely used in robotics, has been tailored to focus on communication and data management. Nodes, the basic components of ROS~2, are small execution units that handle specific functions or processes within the broader robotic system. Communication between nodes occurs via buses called topics that define specific data types and messages. This data exchange employs a publish and subscribe mechanism. Nodes that publish send data to specified topics and nodes that subscribe receive data from these topics. For example, the process of publishing data from the camera and its subsequent subscription by other nodes that require this information is illustrated in Fig.~\ref{fig:pubsub_model}. This flexible communication structure of ROS~2 allows for the efficient integration of various parts of a complex robotic system, facilitating a high degree of automation and cooperative behavior.

\subsection{Many-core Processors}
\label{section:many-core}
\begin{figure}[tb]
　\centering
    \includegraphics[width=1.0\linewidth]{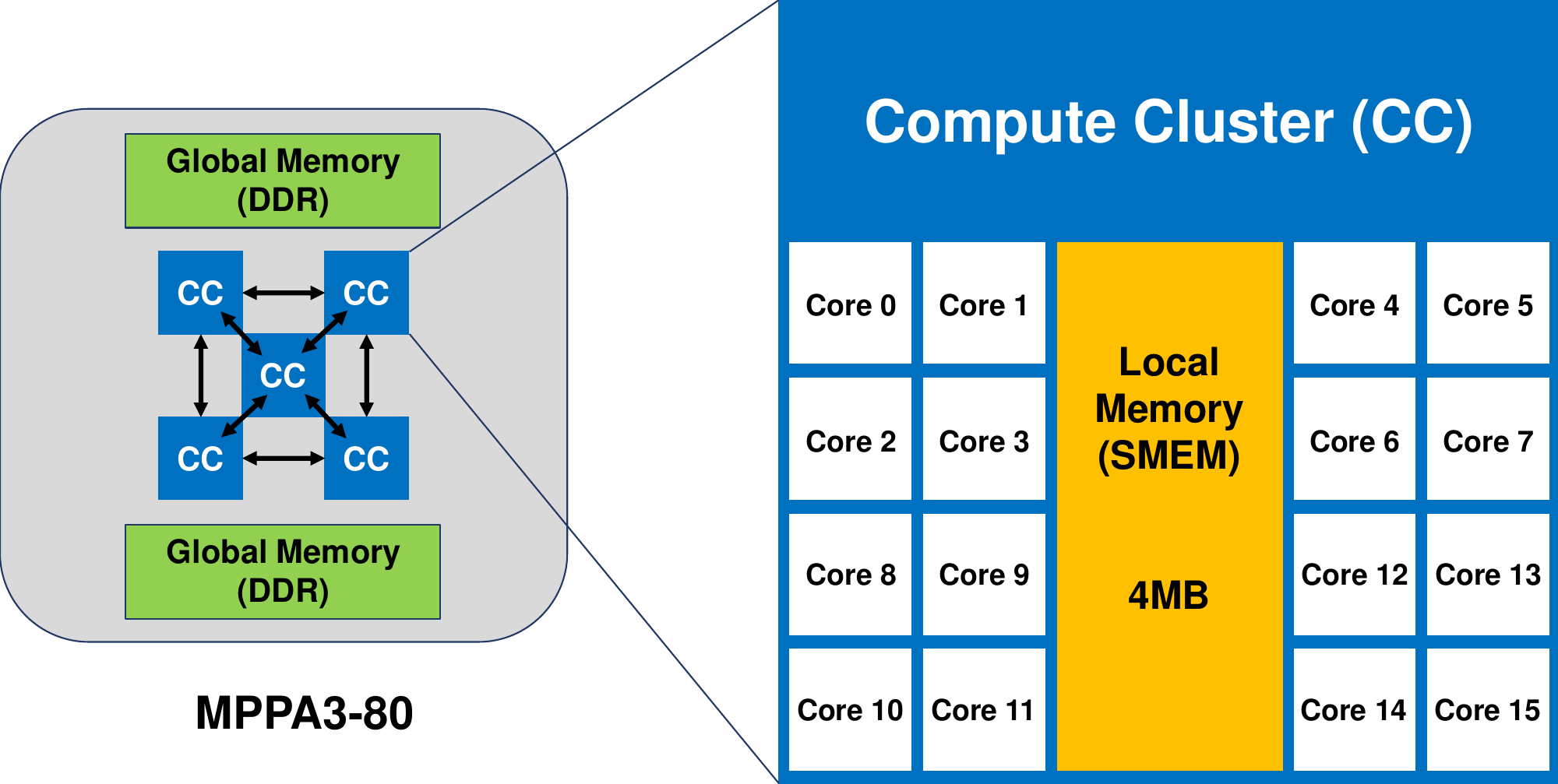}
    \caption{MPPA3-80 Coolidge.}
    \label{Coolidge}
\end{figure}

\begin{figure}[t]
    \begin{center}
        \includegraphics[width=1.0\linewidth]{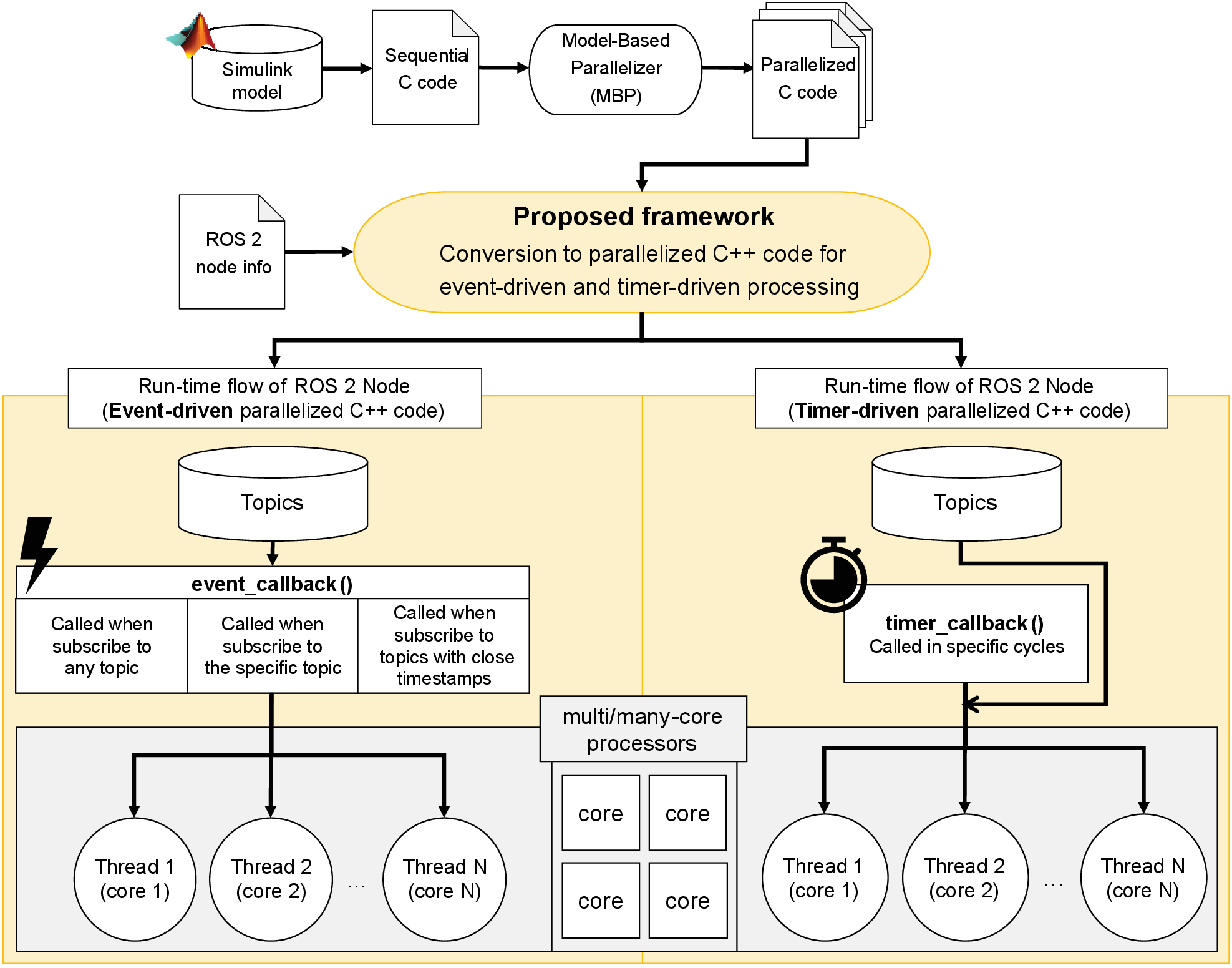}
        \caption{Overview of the proposed framework.}
        \label{fig:framework}
    \end{center}
\end{figure}
  
Many-core processors consist of more cores than multi-core processors. They are characterized by high-performance processing capability and low power consumption. Many-core processors used in this study is Kalray MPPA3-80 Coolidge as shown in Fig.~\ref{Coolidge}, provided by Kalray, which consists of 80 cores divided into five compute clusters (CC). The memory is shared among the CCs via NoC (network-on-chip) communication~\cite{Response}. Addressing challenges related to parallelism and cache coherency is imperative when adopting an existing operating system on a many-core platform. eMCOS, \textit{a real-time operating system} (RTOS), was selected as the operating system for the study. eMCOS is an RTOS that supports many-core processors such as the Kalray MPPA3-80 Coolidge and provides a minimal programming interface and libraries. The distributed microkernel architecture in eMCOS empowers each core to deliver essential functionality independently, minimizing resource contention and ensuring high throughput. Microkernels within each core manage various functionalities, including message passing, inter-core communication, and thread management.

\section{Proposed Framework}
\label{chapter:proposed framework}

The proposed framework converts the parallelized C code generated by MBP into ROS~2-based parallelized C++ code, as shown in Fig.~\ref{fig:framework}. The inputs of the proposed framework are the parallelized C code generated by MBP and the ROS~2 node information obtained from the model. The output of the proposed framework is either event-driven parallelized C++ code of ROS~2 nodes or timer-driven parallelized C++ code of ROS~2 nodes. This section outlines the conversion process:

\begin{itemize}
    \item The design of the parallelized C code and important considerations when converting to parallelized C++ code are described.
    \item The method for converting to timer-driven and event-driven nodes is explained.
\end{itemize}

\subsection{The Design of The Parallelized C Code}

\begin{figure}[tb]
    \centering
    \begin{lstlisting}
void thread1() {
  // Declare and initialize variables
  while(1){
    // Receive input from external
    // Process
    // Output to external
  }
}
void thread2() {
  // Same as thread1
}
int main() {
  // Create threads
  // Assign processor cores to threads
  // Start threads
  return 0;
}
\end{lstlisting}
    \caption{MBP-generated parallelized C code design for two cores.}
    \label{fig:MBP-design}
\end{figure}

An example of the design of the parallelized C code generated by MBP is shown in Fig.~\ref{fig:MBP-design}. Here, processing is distributed across two cores, with each core handling a separate thread. Initially, variables are declared and initialized in each thread, and channels are created for inter-core communication. Then, processing is performed for each core within a while loop. 

On the other hand, in the Publish/Subscribe model of ROS~2, when a message is published to a topic, it is processed based on that event. This mechanism is asynchronous and does not require waiting for new messages, allowing for non-blocking operations.

\cthree{
MBP cannot directly utilize the Publish and Subscribe blocks of ROS~2 in Simulink. Therefore, an existing method~\cite{ryoETFA2020} of replacing these blocks with other blocks with equivalent functionality, but the data types can only be recognized as the basic data types provided in Simulink. This requires additional input of ROS~2 node information such as topic name, message type and callback function name.
}

\subsection{Timer-Driven Nodes}
\label{section:timer-driven nodes}
\begin{figure}[tb]
  \begin{center}
    \includegraphics[width=0.9\linewidth]{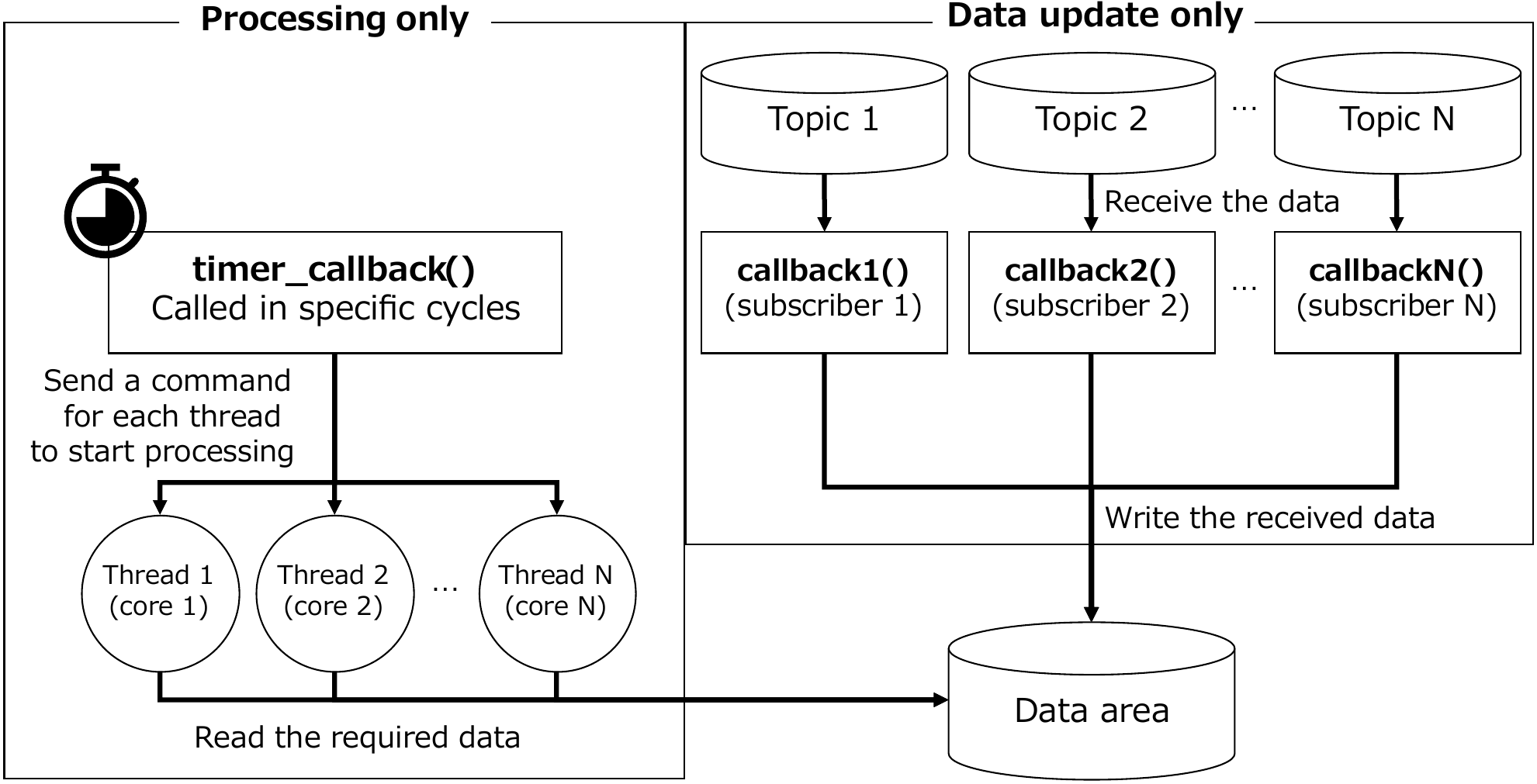}
    \caption{Timer-driven node.}
    \label{fig:timer-driven}
  \end{center}
\end{figure}

\cone{
Timer-driven nodes start processing at regular time intervals. Timer-driven processing is used, for example, for the process of receiving point clouds from LiDAR. The structure is shown in Fig.~\ref{fig:timer-driven}. In this structure, the timer callback function is called at regular time intervals. The ``Processing only'' section on the left-hand side of the system initiates the $timer\_callback()$ function at defined intervals, triggering multiple threads to execute specific data processing tasks concurrently. Meanwhile, the ``Data update only'' section on the right handles the data-driven aspect of the system. Each data topic (from $Topic\;1$ to $Topic\;N$) is subscribed to by the corresponding callback function ($callback1()$ through $callbackN()$) to write to the central data area.
}

\subsection{Event-Driven Nodes}
\label{section:event-driven nodes}
\ctwo{
Event-driven nodes start processing when an event occurs. Event-driven processing is used, for example, for self-position estimation and velocity control. In this paper, we divide event-driven nodes into three patterns: (1) nodes that start processing when data is received from all input topics, (2) nodes that start processing when data is received from a specific input topic, and (3) nodes that start processing when data is received from a specific input topic and the timestamp is close.
}

\subsubsection{Event-Driven Node to Start Processing by All Input Topics}
\begin{figure}[tb]
  \begin{center}
      \includegraphics[width=0.9\linewidth]{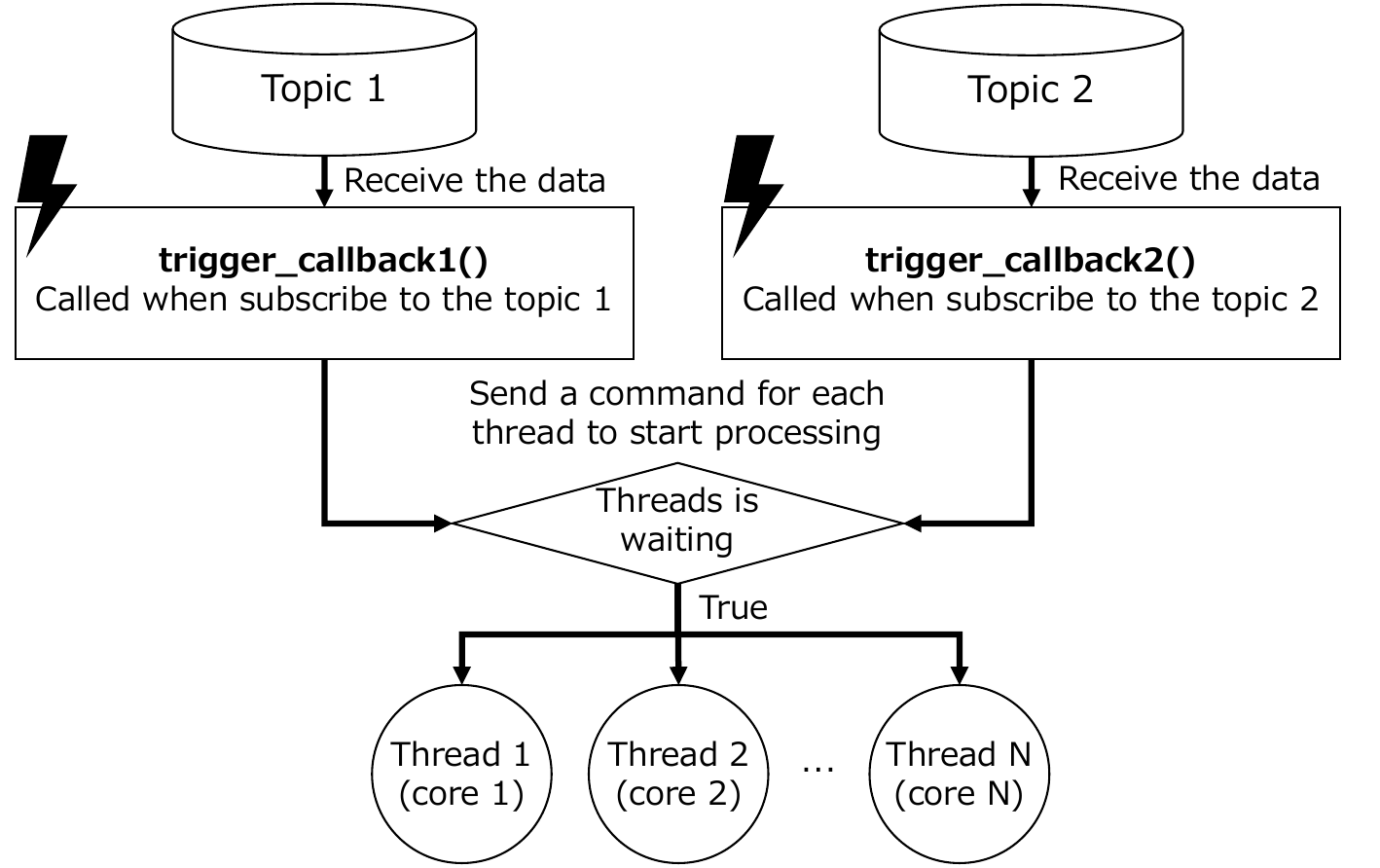}
      \caption{Event-driven node to start processing all input topics.}
      \label{fig:event-driven_all}
  \end{center}
\end{figure}

\ctwo{
An event-driven architecture in which all inputs are treated as events is shown in Fig.~\ref{fig:event-driven_all}. In this structure, when data is received from $Topic\;1$ and $Topic\;2$, $trigger\_callback1()$ and $trigger\_callback2()$ are called, respectively. These callback functions are triggered by the receipt of data from the subscribed topic, and a signal is sent to the threads (from $Thread\;1$ to $Thread\;N$) to start processing. Threads monitor this signal through condition variables and start data processing as soon as they are triggered.
}

\begin{figure}[tb]
    \begin{center}
        \includegraphics[width=0.9\linewidth]{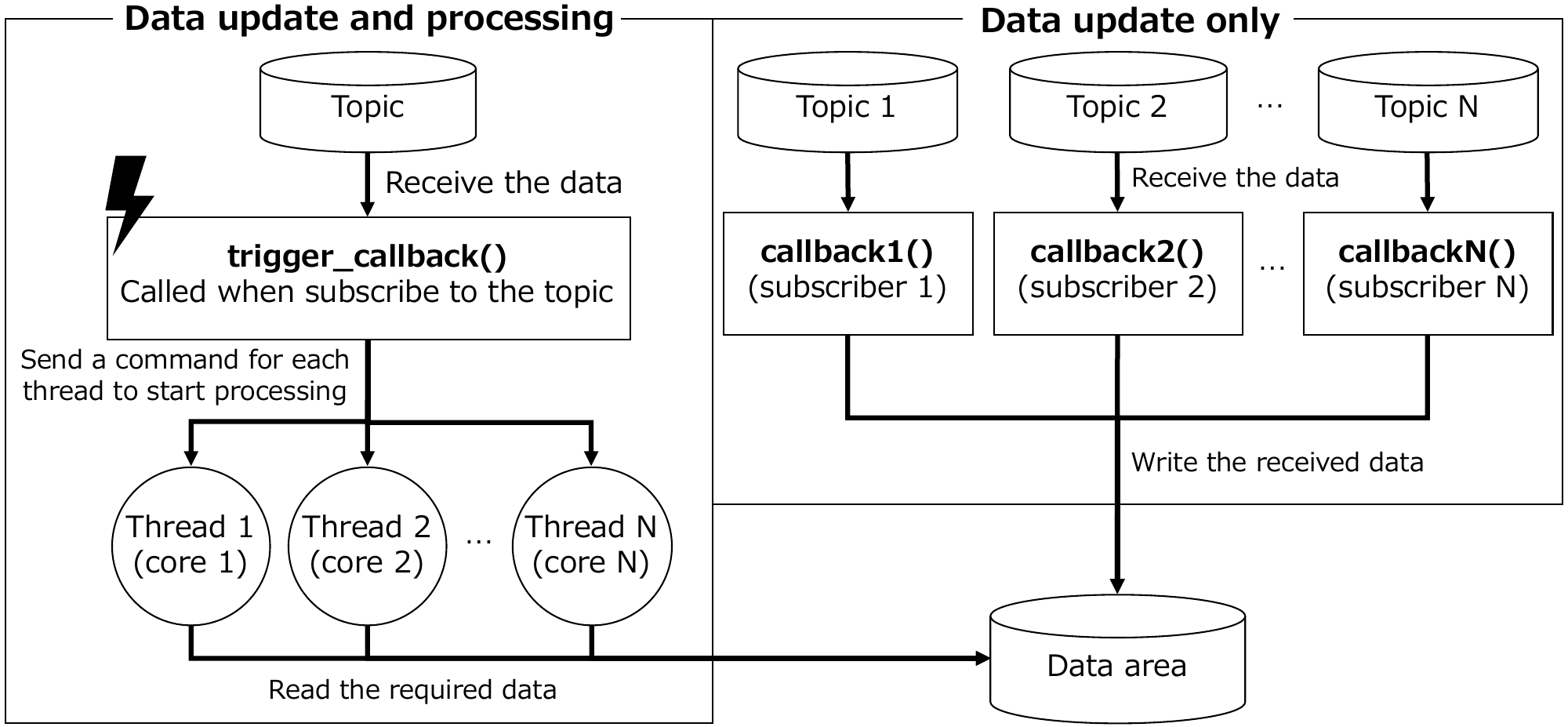}
        \caption{Event-driven node to start processing only specific input topic.}
        \label{fig:event-driven_trigger}
    \end{center}
\end{figure}

\begin{figure}[tb]
    \begin{center}
        \includegraphics[width=0.9\linewidth]{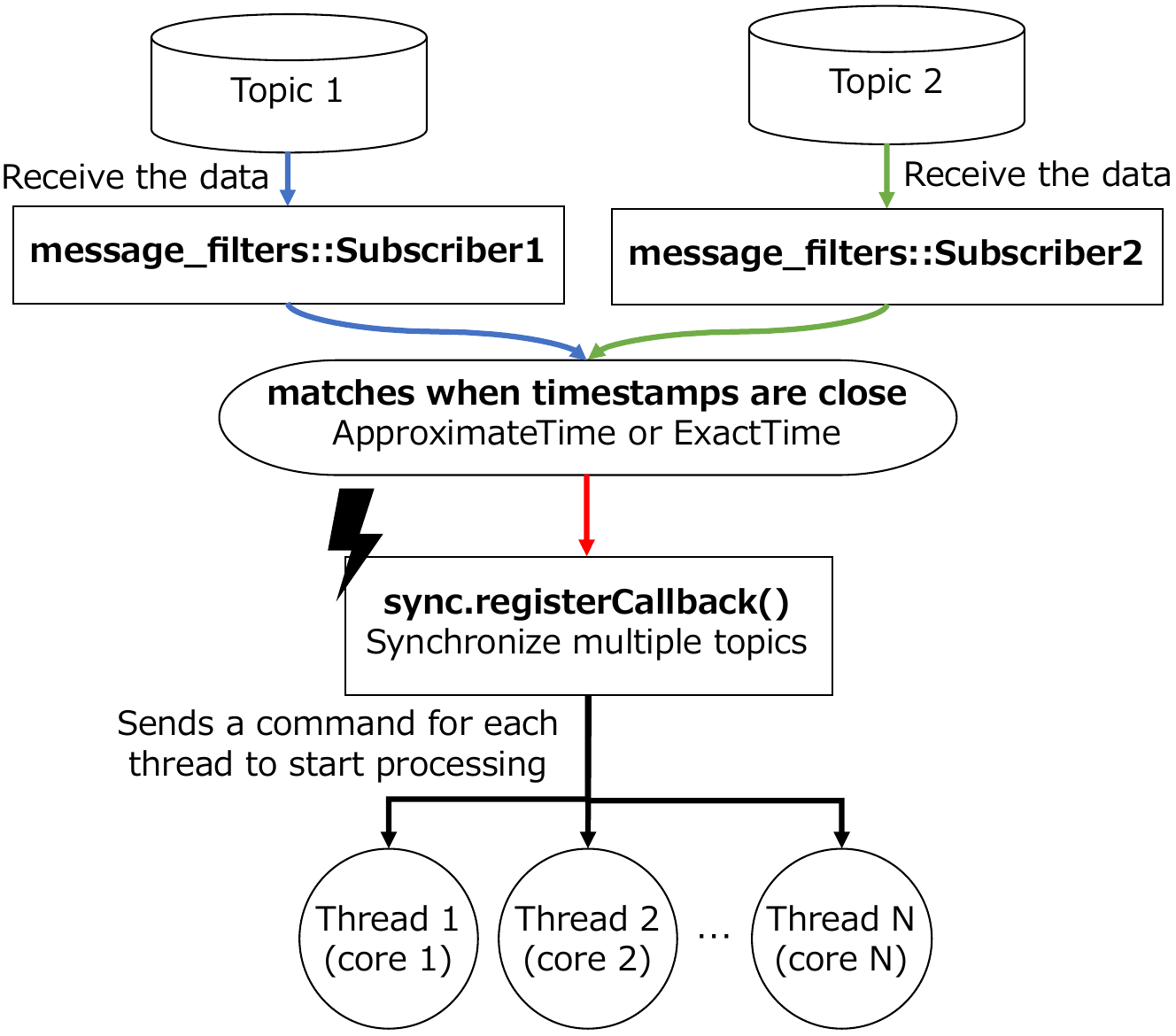}
        \caption{Event-driven node to start processing if timestamp is close.}
        \label{fig:event-driven_mf}
    \end{center}
\end{figure}


\subsubsection{Event-Driven Node to Start Processing Only Specific Input Topic}
\ctwo{
An event-driven architecture in which only specific inputs are treated as events is shown in Fig.~\ref{fig:event-driven_trigger}. In the `Data update and processing' section on the left, a $trigger\_callback()$ is called upon the reception of data from a topic. This trigger function occurs when subscribing to the related topic and is responsible for sending commands to each thread (from $Thread\;1$ to $Thread\;N$) to start data processing. On the right, the ``Data update only'' section shows that data from $Topic\;1$ to $Topic\;N$ is received by their respective callback functions (from $callback1()$ to $callbackN()$). These functions are associated with subscribers and write the received data to the central data area.
}


\subsubsection{Event-Driven Node to Start Processing If Timestamp Is Close}

\ctwo{
An event-driven architecture in which processing starts when the timestamp is close is shown in Fig.~\ref{fig:event-driven_mf}. In this architecture, data from two distinct topics, $Topic\;1$ and $Topic\;2$, are received by \seqsplit{$message\_filters::Subscriber1$} and  \seqsplit{$message\_filters::Subscriber2$}, respectively. These subscribers use the $ApproximateTime$ or $ExactTime$ policy, and data is only processed if both messages match the timestamp. The $sync.registerCallback()$ function is invoked only when the timestamps meet the criteria. This synchronization registration callback function coordinates the timestamps across multiple topics, sending commands to start data processing to each thread from $Thread\;1$ to $Thread\;N$. 
}

\subsection{The Design of The Proposed Parallelized C Code}
\label{The Design of The Proposed Parallelized C Code}
\begin{figure}[tb]
    \centering
    \begin{lstlisting}
void callback1() {
  // Update data
}
void callback2() {
  // Update data
}
void timer_callback() {
  // Use condition variables to notify each thread to start
}
void trigger_callback1() {
  // Update data
  // Use condition variables to notify each thread to start
}
void trigger_callback2() {
  // Update data
  // Use condition variables to notify each thread to start
}
void sync_callback() {
  // Update data
  // Use condition variables to notify each thread to start
}
void thread1() {
  while(ROS 2 node is alive) {
    // Wait in a standby state for notification from callbacks
    // processes
  }
}
void thread2() {
  // Same as thread1
}
int main() {
  rclcpp::init(argc, argv);
  // Create subscribers and publisher
  // Create timer and link with timer_callback
  // Create threads
  // Assign processor cores to threads
  // Start threads
  rclcpp::spin(node);
  rclcpp::shutdown();
  return 0;
}
\end{lstlisting}
    \caption{Comprehensive framework for parallelized C++ code in ROS~2: timer and event-driven nodes on a two-core system.}
    \label{fig:Comprehensive}
\end{figure}
The code of the proposed approach is shown in Fig.~\ref{fig:Comprehensive}, with an example of timer-driven nodes and event-driven nodes with two processor cores and two additional callbacks. The code shows how the proposed approach makes efficient use of resources and optimizes parallel processing.

In the timer-driven nodes, the $timer\_callback()$ function is triggered periodically to notify $thread1()$ and $thread2()$ to start processing. This notification causes the threads to begin processing data. Threads remain in a wait state within the while loop and begin processing when notified; the two callback functions $callback1()$ and $callback2()$ are used to update and process the data received by the node.

In the event-driven nodes that start processing on all input topics, $trigger\_callback1()$ and $trigger\_callback2()$ are defined to update data when data is received from each topic and send a notification to the thread to start processing. For event-driven nodes that only start processing certain input topics, the $trigger\_callback()$ is called when data is received and sends a notification to the thread to start processing. For event-driven nodes that start processing when the timestamp is close, the $sync\_callback()$ updates the data based on the timestamp and sends a notification to the thread to start processing.

The $main()$ function handles initialization of the ROS~2 node, creation of subscribers and publishers, setting timers, assigning processor cores to threads, and starting threads. When threads are created, they are assigned to their respective processor cores. $rclcpp::spin$ starts the event loop and $rclcpp::shutdown$ shuts down the node properly.

\begin{figure}[tb]
  \centering
  \includegraphics[width=1.0\linewidth]{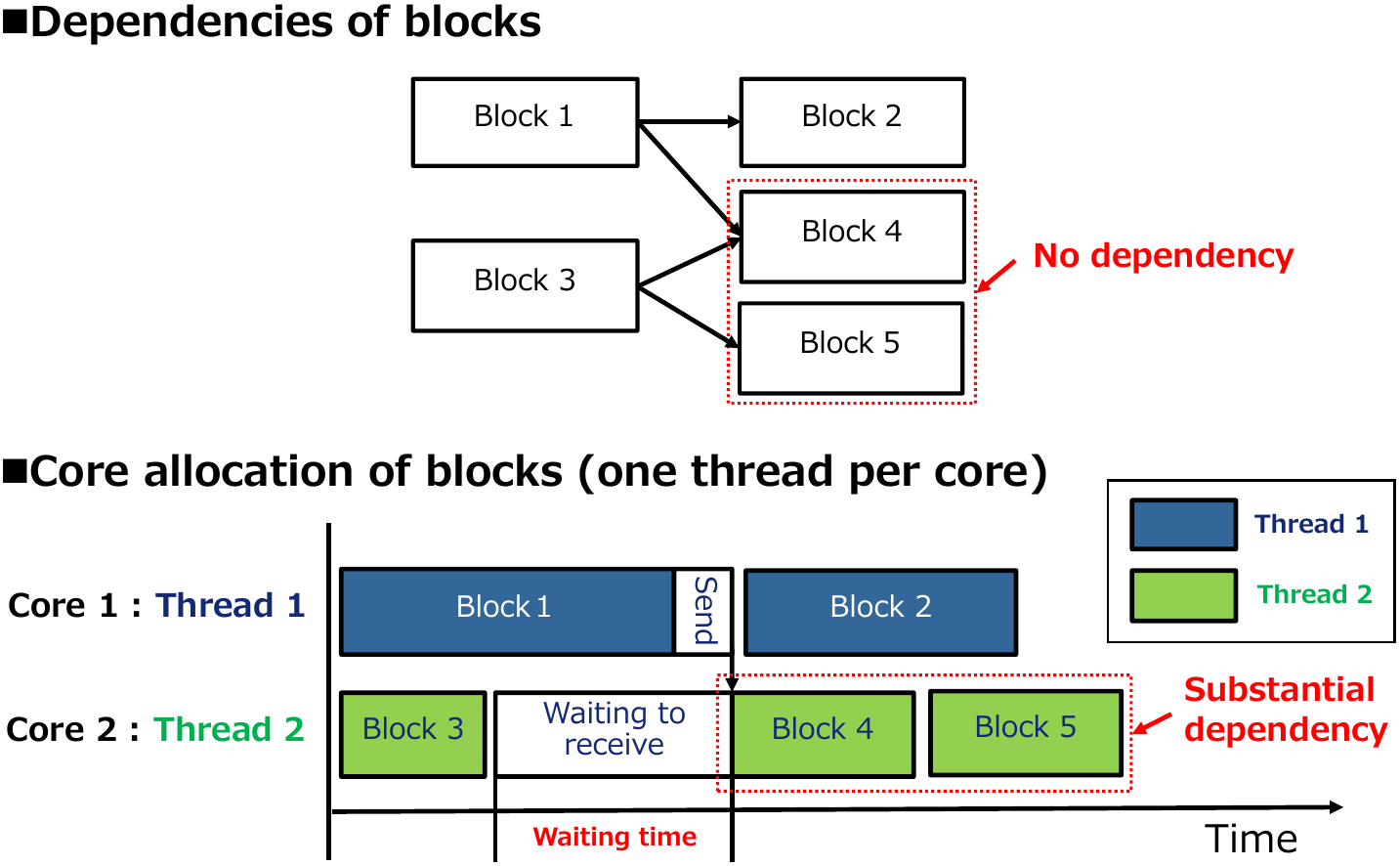}
  \caption{Illustrative example highlighting the issue of prolonged waiting time due to block dependencies and core allocation inefficiencies in existing methods.}
  \label{fig:multithread_existing_en}
\end{figure}

\begin{figure}[tb]
  \centering
  \includegraphics[width=1.0\linewidth]{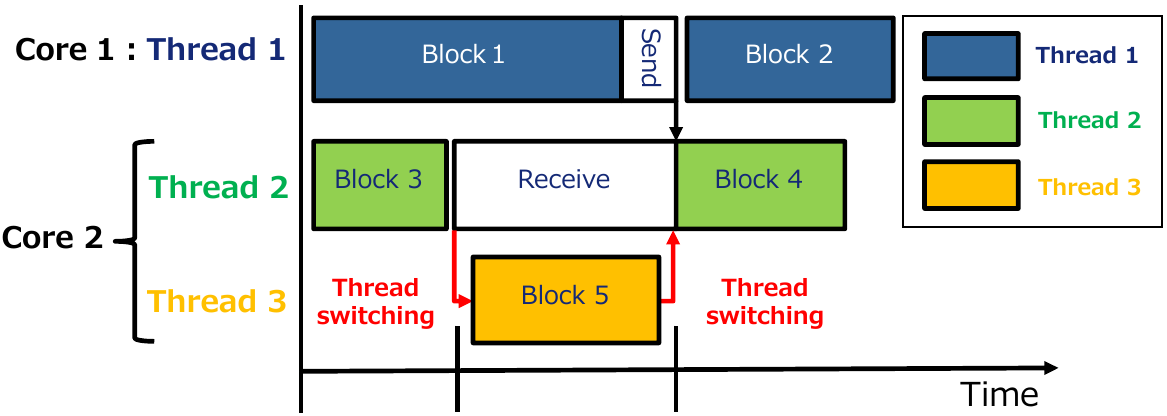}
  \caption{Example demonstrating the efficient use of waiting time through core and thread allocation in the proposed method.}
  \label{fig:multithread_proposed_en}
\end{figure}

\subsection{Parallel Performance Improvement}
\label{section:parallel performance improvement}

This section explains techniques to increase CPU utilization by increasing threads for models that are not highly parallel. A model with high parallelism is characterized by evenly distributing tasks to different cores, with minimal communication between them. On the other hand, a model with low parallelism is one in which tasks cannot be distributed appropriately to different processor cores due to task dependencies and other factors. Furthermore, the more communication between cores, the lower the improvement effect of parallelization.

The parallelization code generated by MBP allocates only one thread per core. Therefore, when a core needs to wait for inter-core communication, it cannot perform any other processing. On the other hand, by assigning multiple threads to one core, it can switch to another thread and continue processing even when it needs to wait for inter-core communication. This results in higher CPU utilization and shorter overall program execution time. A motivation example is shown in Fig.~\ref{fig:multithread_existing_en}. In this case, inter-core communication is performed from $Block\;1$ to $Block\;4$, and $Block\;4$ waits until $Block\;1$ transmits. In addition, $Block\;4$ and $Block\;5$ have no dependencies, while $Block\;5$ must wait for $Block\;4$ to finish. Therefore, $Core\;2$ becomes idle, and CPU utilization decreases. On the other hand, as shown in Fig.~\ref{fig:multithread_proposed_en}, by assigning $Blocks 3$ and $Block\;4$ to $Thread\;1$ and $Block\;5$ to $Thread\;2$, the thread of $Core\;2$ switches until $Block\;1$ transmits. In other words, since $Block\;4$ and $Block\;5$ have no dependencies, $Block\;5$ does not need to wait for $Block\;4$ to finish if the threads are separate. As a result, $Core\;2$ will not be idle, and CPU utilization will increase.

One approach to divide a thread into multiple threads is to use the core allocation of MBP. This approach performs parallelization with a more significant number of processor cores than actually used and then changes the core allocation of the generated threads to the processor cores to be used. For example, using the result of parallelization for four processor cores in MBP, the core allocation is changed to two cores. Thus, the number of threads per core can be increased. This approach differs from the ROS~2 feature MultiThreadedExecutor~\cite{Experimental}, where the executor is able to parallelize nodes, but not a single callback into multiple threads. The proposed method in this paper can solve this issue.

\begin{table}[tb]
  \centering
  \caption{Evaluation environment of POSIX}
  \label{table:evaluation environment}
  \begin{tabularx}{\linewidth}{|X|l|}
  \hline
  Processor & Intel Core i5-12400F @ 2.50 GHz \\
  \hline
  Number of Physical Cores & 6 \\
  \hline
  Memory & 32.0 GB \\
  \hline
  OS & Windows 10 and Ubuntu 22.04 LTS \\
  \hline
  ROS 2 Version & Humble Hawksbill \\
  \hline
  eMBP Version & 2.3.1 \\
  
  \hline
  \end{tabularx}
\end{table}

\subsection{Support for Many-core Platforms}
\label{section:Support for Many-core Platforms}

In MBP, when parallelized code is executed in Coolidge, deadlocks sometimes occur depending on the structure of the model. The reason is that they use a communication method where threads occupy cores. This communication method tries to achieve stable communication by occupying cores, but it cannot cope with deadlocks. Thus, a conversion script has been created to ensure that threads do not occupy cores. After executing the conversion script in the parallelized code, a compiler is applied so that it can be executed on Coolidge.

\section{Evaluation}
\label{chapter:evaluation}

\begin{figure}[tb]
    \centering
  \includegraphics[width=1.0\linewidth]{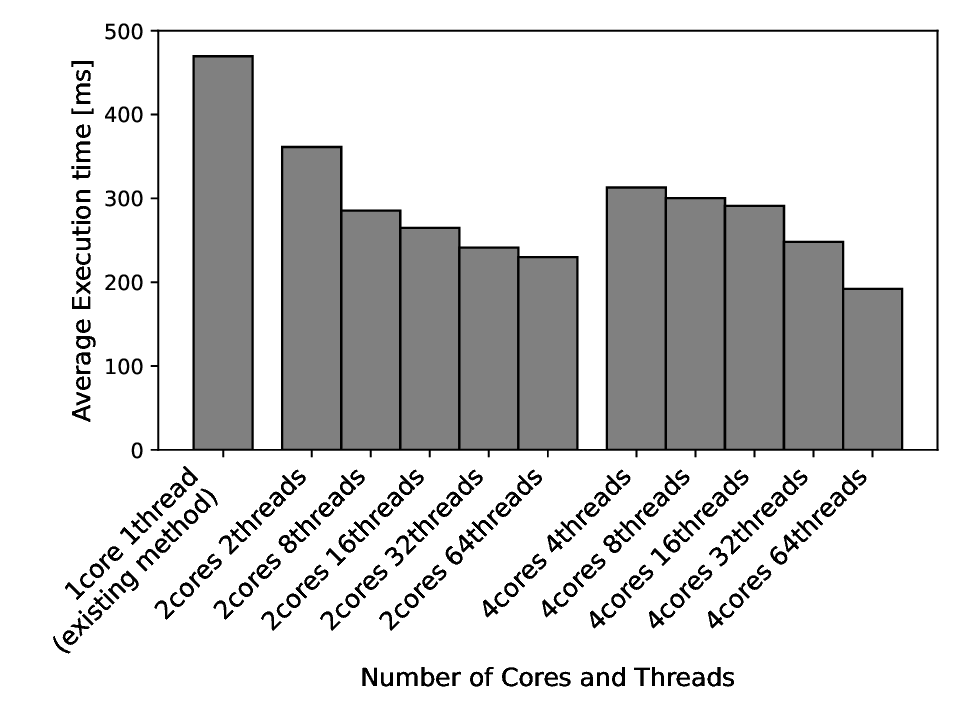}
  \caption{Average execution time for multiple processor cores and threads in POSIX environment.}
  \label{fig:runtime_posix}
\end{figure}
  
\begin{figure}[tb]
  \centering
  \includegraphics[width=1.0\linewidth]{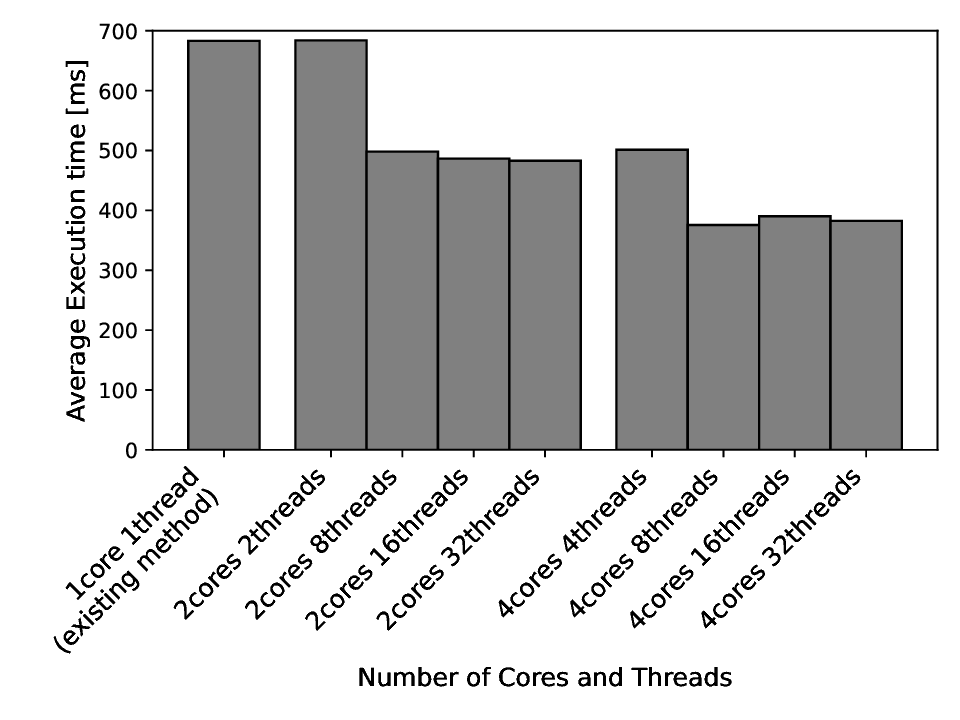}
  \caption{Average execution time for multiple processor cores and threads in the eMCOS environment.}
  \label{fig:runtime_emcos}
\end{figure}

In this section, we evaluate the proposed framework. Two evaluation environments were used. The POSIX environment is shown in Table~\ref{table:evaluation environment}, eMCOS environment is described in Section~\ref{section:many-core}. First, the proposed method was applied to a highly parallel Simulink model in the eMCOS environment. As shown in Fig.~14, the execution time improves as the number of cores increases. The reason for this is that this Simulink model is a model with low inter-core communication and low communication overhead. Next, a random model was evaluated. The random model was generated stochastically and automatically using the capabilities of an existing method, SLForge~\cite{chowdhury2020slemi}. To confirm effectiveness of parallelization with the proposed framework, we measured the average execution time of the parallelized programs. The execution time is the time from the start of processing to the end of processing, i.e., from data submission to publication. The measurement was repeated 1,000 times, and the median 80\% average of the obtained values was used as the final average execution time.

\subsection{Basic Evaluation of Random Simulink Model}
\label{section:basic evaluation of random simulink model}

The average execution time of random models parallelized on 1, 2 and 4 processor cores by the proposed framework in the POSIX environment was measured. The results are shown in Fig.~\ref{fig:runtime_posix}. As the number of processor cores increases, the reduction in execution time is evident. However, since the model is not highly parallel, the reduction in execution time as the number of processor cores increases is limited. We tried the event-driven and timer-driven patterns and found that both patterns had about the same execution time. This indicates that parallelization is appropriate for both patterns.

The execution results for the eMCOS environment are shown in Fig.~\ref{fig:runtime_emcos}. As the number of cores increases, the overall execution time decreases. The difference between one core and two cores is the presence of channels. For two or more cores, a communication channel is created for inter-core communication. The random model used in the evaluation is a model with much inter-core communication. Therefore, the difference in execution time between one core and two cores is considered to be small due to the increased communication overhead. For the 2-core and 4-core models, the improvement in performance due to parallelization exceeded the communication overhead, which is believed to have reduced the execution time. For the same number of cores and threads, the execution time differs between the POSIX and eMCOS environments. The reason for this is the difference in frequency: 2.5~GHz in the POSIX environment and 1.0~GHz in the eMCOS environment.

\subsection{Evaluation of Multi-threaded Random Simulink Model}
\label{section:evaluation with random simulink model}

The execution time was measured when the number of threads was increased in a POSIX environment. The results are shown in Fig.~\ref{fig:runtime_posix}. The number of threads allocated to the two processor cores was increased from 8 to 64. The maximum number of threads was set to 64 because of MBP constraints in the random model. Here, the scheduling policy for each thread was set to $SCHED\_FIFO$. Only the threads of the starting process were set to a lower priority. Splitting a large number of threads significantly reduces execution time. This is because the increased number of threads allows another process to be performed while waiting for communication, thus improving CPU utilization.

The execution results for the eMCOS environment are shown in Fig.~\ref{fig:runtime_emcos}. The results show the execution time when the number of threads is increased from 8 to 32 for the 2-core and 4-core cases. The reason for setting the maximum number of threads to 32 is that eMCOS is limited to 32 threads per core. As in the POSIX environment, the higher the number of threads, the shorter the execution time for the same number of cores. The reason for this is thought to be that the amount of processing per core has increased. In the 4-core case, it can be seen that the execution time increases as the number of threads increases from 8 to 16. The reason for this is thought to be that as the number of threads increases, the cost of context switches to switch threads also increases.

\begin{table*}
  \caption{Comparison between proposed framework and previous study}
  \label{table:comparison}
  \centering
  \scriptsize
 {
  \begin{tabular}{|l|c|c|c|c|c|c|c|}
    \hline
    & \textbf{Embedded} & \textbf{Autonomous}& \textbf{MATLAB/} & \textbf{Multi-core} & \textbf{ROS} & \textbf{ROS 2} & \textbf{Event-driven and} \\
    & \textbf{system} &  \textbf{driving} & \textbf{Simulink} & \textbf{processors}  &  & & \textbf{timer-driven} \\
    \hline
    Scilab/Xcos, Scade \cite{Reder2019WCET} \cite{Didier2019Correct} & \(\checkmark\) & &  & \(\checkmark\) & & & \\ \hline
    CoCoSim \cite{Bourbouh2020CoCosim} & \(\checkmark\) &  &  \(\checkmark\) &  & & & \\ \hline
    \multirow{2}{*}{\begin{tabular}[l]{@{}l@{}}Parallelization of Simulink model \\ \cite{Tunacali2016Automatic}\cite{Bansal2018optimal}\cite{Morelli2015SysML} \cite{zhong2019model}\end{tabular}} & \multirow{2}{*}{\(\checkmark\)} &  & \multirow{2}{*}{\(\checkmark\)} & \multirow{2}{*}{\(\checkmark\)} & & & \\ 
    & & & & & & & \\
    \hline
    Modular MBD \cite{david2022modular} & \(\checkmark\) &  & \(\checkmark\) & & \(\checkmark\) & & \\ \hline
    ETFA 2020 \cite{ryoETFA2020}  & \(\checkmark\) & \(\checkmark\) & \(\checkmark\) & \(\checkmark\) & \(\checkmark\) & & \\ \hline
    EUC 2023 \cite{jyakumi2023euc}  & \(\checkmark\) & \(\checkmark\) & \(\checkmark\) & \(\checkmark\) &  & \(\checkmark\) & \\ \hline
    Proposed framework & \(\checkmark\) & \(\checkmark\) & \(\checkmark\) & \(\checkmark\) & & \(\checkmark\) & \(\checkmark\) \\
    \hline
  \end{tabular}
  }
\end{table*}

\section{Related Work}
\label{chapter:related work}

This section presents research on parallelization using model-based development. Initially, the section covers research on parallelization from MBD tools other than MATLAB/Simulink, followed by an exploration of parallelization from Simulink models. Finally, studies targeting Simulink models in ROS or ROS~2 are presented.

The other MBD tools besides Simulink are Scilab/Xcos~\cite{Scilab/Xcos} and ANSYS SCADE Suite~\cite{SCADE} Xcos optimizes parallel programs using automatic parallelization with worst-case execution time (WCET) and interference analysis~\cite{Reder2019WCET} to achieve low interference and high predictability. Real-time scheduling algorithms~\cite{Didier2019Correct} utilize the SCADE language to allow predictable execution by carefully tuning the analysis, mapping, and code generation phases. These phases account for resource access interference from multi-core execution. CoCoSim~\cite{Bourbouh2020CoCosim}, a framework for converting Simulink models to synchronous programming languages, provides two capabilities for the analysis and code generation of multi-rate Simulink models. However, support for multi-core processors is under consideration.

Research has also been done to make effective use of multi-core processors in the Simulink environment. A heuristic method~\cite{Tunacali2016Automatic} partially determines the execution order when independent blocks are mapped to the same core. This approach can return a good solution with reasonable execution time, even for models with a large number of blocks. On the other hand, they only present a simple example for multi-rate models and state that the proposed method is not practical for multi-rate models. An optimization method~\cite{Bansal2018optimal} addresses the problem of optimizing software implementations of Simulink models under split-preemptive fixed-priority scheduling on multi-core platforms. This method can optimize the software implementation of Simulink models by assigning task priorities, assigning task offsets and adding rate transition blocks.

Research has also been done to integrate Simulink with other tools. A design methodology for robotic systems~\cite{Morelli2015SysML} integrates Simulink to define the functionality and a SysML model to define the mapping to the execution platform. Platform modeling prepares the analysis and evaluation of architectural design choices and the analysis of computation and communication delays caused by architectural geometry and functionality. A method~\cite{zhong2019model} efficiently parallelizes Simulink models on multi-core CPUs and GPUs. In this method, the Simulink model is first converted into a Directed Acyclic Graph (DAG). Integer linear programming (ILP) is then used to allocate tasks to CPU cores and GPUs, taking into account task execution time and communication time. This ensures that inter-task dependencies and load are managed efficiently, reducing the execution time of image processing algorithms.

Research on developing ROS or ROS~2 applications with MBD has also been conducted recently. A method~\cite{ryoETFA2020} generates parallelized C++ code for ROS from code generated by MBP. The proposed method avoids the restriction of MBP by deleting the blocks for ROS temporarily, and after parallelization by MBP, the temporarily deleted blocks for ROS are restored manually to parallelize the Simulink model for ROS. For ROS~2-based Simulink models, the method of task and data parallelization~\cite{jyakumi2023euc} replaces ROS~2 blocks automatically with existing blocks. This approach avoids the problem that MBP cannot use the latest blocks containing blocks for ROS~2. However, since the code generated by MBP is C code, it must be properly converted manually for use as a ROS~2 node.

The characteristics and comparisons of the proposed framework and related work are summarized in Table~\ref{table:comparison}. Existing related work has various parallelization approaches for model-based development. However, no existing work supports model-based parallelization within ROS~2 while also accommodating event-driven and timer-driven nodes.


\section{Conclusion}
\label{chapter:conclusion}

In this paper, we proposed a framework to generate parallelized C++ code for ROS~2 from Simulink models automatically, enabling efficient parallel processing. The key point of the proposed method was that it can handle both event-driven nodes, which start processing immediately upon input, and timer-driven nodes, which start processing at regular intervals. In addition, it enabled automatic parallelization of ROS~2-based Simulink models with multiple inputs, which has been difficult with conventional methods. Evaluation results showed that the proposed framework can reduce execution time. The evaluation results also showed that the execution time can be further reduced by increasing the number of threads per core. Since real-time performance is important for autonomous driving systems, this automatic parallelization is expected to make a significant contribution to improving development efficiency and reliability.

One of the future challenges for the proposed framework is to consider parallelization in multiple models first. Although the current framework focuses on the parallelization of a single model, further performance improvement can be expected by optimizing parallelization in multiple interconnected models. In addition, there can be other useful patterns, and addressing them would increase the flexibility of the framework.

\bibliographystyle{IEEEtran}
\bibliography{bstcontrol,reference}

\end{document}